\def\epsilonF{\epsilon_{\text{F}}}
\def\kF{k_{\text{F}}}

\def\me{m_{\text{e}}}

\def\Tc{T_{\text{c}}}
\def\pc{p_{\,\text{c}}}

\def\Hctwo{{H_{\text{c2}}}}

\def\be{\begin{equation}}
\def\ee{\end{equation}}
\def\bea{\begin{eqnarray}}
\def\eea{\end{eqnarray}}
\def\bse{\begin{subequations}}
\def\ese{\end{subequations}}

\documentclass[prl,twocolumn,showpacs,preprintnumbers,amsmath,amssymb]{revtex4}
\usepackage{graphicx}% Include figure files
\usepackage{dcolumn}% Align table columns on decimal point
\usepackage{bm}% bold math
\begin{document}
%\preprint{NSF-KITP-05-31}
\title{Columnar Fluctuations as a Source of Non-Fermi-Liquid
Behavior in Weak Metallic Magnets}
\author{T.R. Kirkpatrick$^{1}$ and D. Belitz$^{2}$}
\affiliation{$^{1}$Institute for Physical Science and Technology and Department
                   of Physics, University of Maryland, College Park, MD 20742\\
         $^{2}$Department of Physics and Theoretical Science Institute, University
                of Oregon, Eugene, OR 97403}
\date{\today}
\begin{abstract}
It is shown that columnar fluctuations, in conjunction with weak quenched
disorder, lead to a $T^{3/2}$ temperature dependence of the electrical
resistivity. This is proposed as an explanation of the observed
non-Fermi-liquid behavior in the helimagnet MnSi, with one possible realization
of the columnar fluctuations provided by skyrmion lines that have independently
been proposed to be present in this material.
%
% 410 characters
\end{abstract}


\maketitle

\begin{figure}[b,h]
\vskip -0mm
\includegraphics[width=8.5cm]{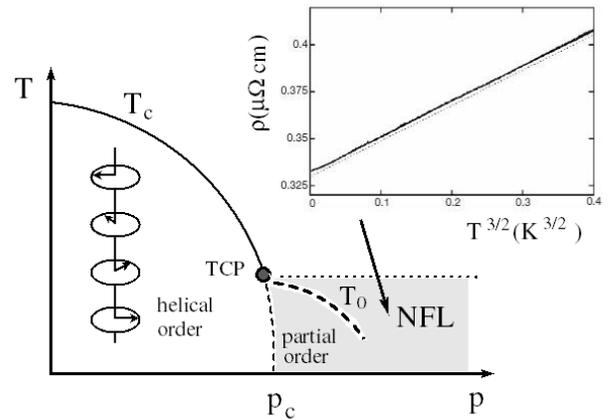}
\caption{Schematic phase diagram of MnSi in the $T$-$p$ plane showing the
 helical phase, the partial-order region, and
 the NFL region. The transition to the
 helical phase is second order (solid line) above a
 tricritical temperature $T_{\text{TCP}}\approx 10\,\text{K}$, and first order
 (dashed line) below. Non-Fermi-liquid (NFL) behavior characterized by an electric
 resistivity $\rho \propto T^{3/2}$ is observed in the paramagnetic phase below
 about $10\,\text{K}$ up to a pressure of at least $2\pc$. The boundary of the NFL
 region is not sharp. The data in the inset are from Ref.\
 \onlinecite{Pfleiderer_Julian_Lonzarich_2001}.}
\label{fig:1}
\end{figure}

\begin{figure}[t]
\vskip -0mm
\includegraphics[width=8.5cm]{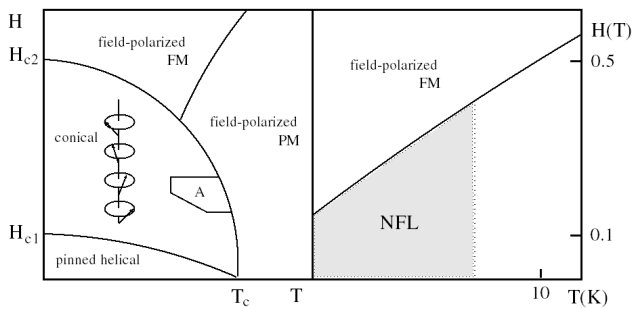}
\caption{Schematic phase diagram of MnSi in the $H$-$T$ plane at $p=0$ (left
 panel) and $p\agt\pc$ (right panel). At $p=0$ successive phase transitions
 lead from a pinned helical phase to a conical phase to a field polarized
 phase. The $A$-phase has been proposed in Ref.\ \onlinecite{Muhlbauer_et_al_2009}
 to be characterized by the columnar skyrmion texture shown in Fig.\ \ref{fig:3}(b).
 For $p$ just above $\pc$ the NFL behavior shown in Fig.\ \ref{fig:1}
 persists for $H\neq 0$ up to to the ferromagnetic
 boundary \cite{Pfleiderer_Julian_Lonzarich_2001}.}
\label{fig:2}
\end{figure}

Weak metallic ferromagnets, which display long-range order only at temperatures
on the order of $10\,{\text K}$, provide a rich testing ground for fundamental
aspects of condensed-matter physics. A prototypical example is MnSi, which
displays long-range magnetic order below a critical temperature $\Tc \approx
29\,{\text K}$ \cite{Ishikawa_et_al_1976}. This low $\Tc$ implies that the (in
this case ferromagnetic) exchange interaction between the electrons has been
renormalized down from the atomic scale ($\approx 10^5\,{\text K}$) by a factor
of almost $10^4$, and thus has to compete with spin-spin interactions of
relativistic nature that are negligible in magnets with higher values of $\Tc$.
In MnSi, a prominent role is played by the Dzyaloshinsky-Moriya (DM)
interaction \cite{Dzyaloshinsky_1958, Moriya_1960}, which leads to a term ${\bm
M}\cdot({\bm\nabla}\times{\bm M})$ in the free energy density, with ${\bm M}$
the magnetization. Such a term can only occur in systems that are not inversion
symmetric, such as MnSi which crystallizes in the non-centrosymmetric B20
structure, and leads to a helical ground state in the ordered phase, rather
than a ferromagnetic one \cite{Bak_Jensen_1980}. The DM interaction derives
from the spin-orbit interaction, which is weak on the atomic energy scale. As a
result, the pitch wave number $q$ of the helix is small compared to the inverse
atomic length scale; in MnSi, $2\pi/q \approx 180\,\AA$. Crystal-field effects
weakly pin the helix wave vector in the $\langle 1,1,1\rangle$ directions. Upon
the application of hydrostatic pressure, $T_{\text c}$ decreases until it
vanishes at a critical pressure $\pc\approx 14.6\,{\text{kbar}}$
\cite{Pfleiderer_et_al_1997}, see Fig.\ \ref{fig:1}.
Partial magnetic order has been observed in the paramagnetic phase below a
temperature $T_0$ \cite{Pfleiderer_et_al_2004, Uemura_et_al_2007}. In a larger
region within the paramagnetic phase, below a temperature of about
$10\,\text{K}$ and at pressures up to about $2\pc$, the electrical resistivity
$\rho$ displays a $T^{3/2}$ temperature dependence over almost three decades
from a few mK to almost $10\,{\text K}$
\cite{Pfleiderer_Julian_Lonzarich_2001}. This is in striking contrast to
Fermi-liquid theory, which predicts a $T^2$-dependence of $\rho$ at low
temperatures. The boundary of the partial order and the boundary of the
non-Fermi-liquid (NFL) behavior meet the helical phase boundary at roughly the
location of a tricritical point ($T_{\text{TCP}} \approx 10\,\text{K}$) that
separates a line of second-order or very weakly first-order transitions at
higher temperatures from a line of strongly first-order transitions at lower
ones \cite{Pfleiderer_et_al_1997}. In a magnetic field $H$ the magnetization
acquires a homogeneous component in addition to the helix, which leads to the
formation of a conical phase \cite{Ishikawa_et_al_1977}. The amplitude of the
helix vanishes at a critical field $\Hctwo$, above which one has a non-chiral
field-polarized state. The NFL behavior in the disordered phase persists for
$H>0$ up to the boundary of the field-polarized ferromagnetic state
\cite{Pfleiderer_Julian_Lonzarich_2001}. In MnSi, the relevant field scale is
on the order of $0.1 - 0.5\,\text{T}$, see Fig.\ \ref{fig:2}. In an
intermediate field region within the ordered phase near $\Tc$ one observes the
so-called A-phase, which was believed to represent a single helix perpendicular
to the field direction \cite{Grigoriev_et_al_2006a}, but more recently has been
interpreted in terms of spin textures known as skyrmions
\cite{Muhlbauer_et_al_2009}. All of these observations are summarized in Figs.\
\ref{fig:1}, \ref{fig:2}.

A very interesting aspect of helical magnets is that they provide electronic
analogs of liquid crystals. The helical phase is analogous to the ordered phase
in cholesteric liquid crystals, and the helical Goldstone modes are very
similar to the ones found in both cholesteric and smectic liquid crystals
\cite{Belitz_Kirkpatrick_Rosch_2006a}. The partial magnetic order in the
paramagnetic phase is reminiscent of order found in the blue phases of
cholesterics \cite{Pfleiderer_et_al_2004}, and various analogs of blue-phase
physics have been invoked to explain the line labelled $T_0$ in Fig.\
\ref{fig:1} \cite{Tewari_Belitz_Kirkpatrick_2006, Binz_Vishvanath_Aji_2006,
Rossler_Bogdanov_Pfleiderer_2006, Fischer_Shah_Rosch_2008}. These analogies
ultimately break down due to lattice effects in solids at very small energy
scales; how small depends on the system. Order in the spin sector (e.g.,
helical magnets) is advantageous in this respect compared to order in the
charge sector (e.g., density stripe phases \cite{Kivelson_et_al_2003}), since
the electronic spin couples to the ionic lattice only via the spin-orbit
interaction, which is much weaker than the Coulomb interaction.

The NFL behavior observed in MnSi is extremely remarkable. Fermi-liquid theory
is very general, and there are not many established examples of it breaking
down. One is one-dimensional systems, where the concept of a Fermi liquid must
be replaced with that of a Luttinger liquid \cite{Schulz_1995}. In higher
dimensions, NFL behavior is usually associated with quantum critical points
\cite{Sachdev_1999}. If critical soft modes couple to the electron density, NFL
behavior can result. For instance, the NFL linear $T$-dependence of the
resistivity in the normal phase of high-$T_{\text{c}}$ superconductors has been
proposed to be due to a hidden quantum critical point
\cite{Lee_Nagaosa_Wen_2006}. In the case of MnSi this is not a viable
explanation. Not only is the transition at low temperatures first order and
hence there are no critical soft modes, but also the NFL region extends to
pressures much too far from the critical pressure $\pc$ for the system to be in
a critical regime. (There is a quantum critical point at nonzero field
\cite{Belitz_Kirkpatrick_Rollbuehler_2005} that one might invoke , but the
second objection still applies.) Still, soft modes coupling to the relevant
degrees of freedom are the only known mechanism for producing NFL behavior, and
hence an obvious question is the possible existence of generic soft modes that
exist in an entire region of the phase diagram
\cite{Belitz_Kirkpatrick_Vojta_2005}. They can arise from, (1) a spontaneously
broken continuous symmetry, which results in Goldstone modes, or, (2)
conservation laws, or, (3) gauge symmetries. Examples are, (1) magnons, (2)
first sound in fluids or solids, and (3) the massless photon. Simple
considerations show that even if one postulates the existence of a generic soft
mode, it must have unusual properties in order for a $T^{3/2}$ behavior of the
resistivity to result.

We now show that the observed NFL behavior in MnSi can be explained by columnar
spin textures and their fluctuations, which have an unusual dispersion
relation. In addition, weak quenched disorder is necessary to explain the
observations. To motivate this suggestion we stress again that helical magnets
are in many respects electronic analogs of liquid crystals. Columnar phases are
known to exist in the latter \cite{DeGennes_Prost_1993}, and it is natural to
expect them in helical magnets as well. Indeed, a particular realization of
columnar order, namely, spin textures known as skyrmions, have been proposed to
exist in MnSi both in a magnetic field (in the A-phase
\cite{Muhlbauer_et_al_2009}), and in zero magnetic field
\cite{Rossler_Bogdanov_Pfleiderer_2006}. For our purposes the particular
realization of the columnar order is not important, and neither is the
existence of long-range order in the columnar phase; our conclusions remain
valid as long as columns exist.

\begin{figure}[t]
\vskip -0mm
\includegraphics[width=8.5cm]{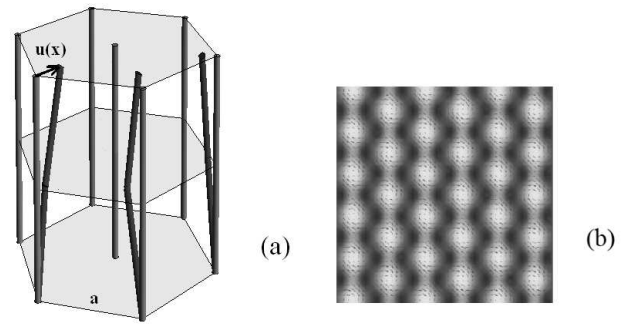}
%\includegraphics[width=5.0cm]{fig_3a.eps}
%\vskip 5mm
%\includegraphics[width=4.0cm]{fig_3b.eps}
\vskip 0mm
\caption{(a) Schematic depiction of columnar order forming a hexagonal lattice
 with lattice constant $a$, and (b) a particular realization of columnar order by
 means of a skyrmion spin texture. The columns fluctuate about their equilibrium
 positions, with ${\bm u}({\bm x})$ the
 displacement vector. In a skyrmion texture, the spins point
 down at the center of each column, and up along the boundary of each cell
 surrounding a column. These textures are similar to the double-twist cylinder
 structures that are believed to be important for the formation of the blue
 phases in liquid crystals mentioned in the text. They can be realized
 by three co-planar helices whose pitch vectors are perpendicular to the column
 axis \cite{Muhlbauer_et_al_2009}. In (b), the light (dark) shading corresponds
 to spins pointing up (down), and the arrows denote the spins projected onto the
 $x$-$y$ plane.}
\label{fig:3}
\end{figure}
Let us assume columnar order forming a two-dimensional hexagonal lattice with
lattice constant $a$, as shown in Fig.\ \ref{fig:3}. Let the columns point in
$z$-direction, and we denote the direction perpendicular to $z$ by $\perp$.
Elasticity theory \cite{Landau_Lifshitz_VII_1986} shows that fluctuations with
a wave vector ${\bm k}$ in the $\perp$ direction will have the dispersion of
ordinary phonons, i.e., the frequency squared will scale linearly with ${\bm
k}_{\perp}^2$. However, since the energy of the lattice cannot change under a
uniform rotation of the columns, the frequency cannot depend on $k_z^2$, and
the lowest power of $k_z$ allowed is $k_z^4$, which can be seen as follows.

Consider an equilibrium state that is a hexagonal lattice of lines, Fig.\
\ref{fig:3}(a), described by two-dimensional lattice vectors ${\bm R}_n = (X_n
, Y_n)$. The fluctuations of the lattice can be expressed in terms of a
two-dimensional displacement field ${\bm u}({\bm x})$, with the fluctuating
lines given by
\be
{\bm r}_n({\bm z}) = (X_n + u_x({\bm R}_n,z),Y_n + u_y({\bm R}_n,z),z).
\label{eq:3}
\ee
The free energy $F$ associated with the harmonic fluctuations of the lattice
can be expressed in terms of the derivatives of ${\bm u}$ as follows
\cite{DeGennes_Prost_1993}
\bea
F &=& \frac{1}{2} \int d{\bm x}\,\biggl[ B\left(\partial_x u_x + \partial_y
u_y\right)^2 + C\Bigl[\left(\partial_x u_x - \partial_y u_y\right)^2
\nonumber\\
&& \hskip -10pt + \left(\partial_x u_y + \partial_y u_x\right)^2\Bigr] +
K\left[\left(\partial_z^2 u_x\right)^2 + \left(\partial_z^2
u_y\right)^2\right]\biggr],
\label{eq:4}
\eea
where $B$, $C$, and $K$ are elastic constants. The absence of a tilt modulus
proper, which would multiply a term quadratic in $\partial_z {\bm u}$, is due
to rotational invariance. The diagonalization of the quadratic form given by
Eq.\ (\ref{eq:4}) yields two eigenvalues which, in Fourier space, have the form
\be
\lambda_{1,2} = \alpha_{1,2}\, {\bm k}_{\perp}^2 + K\,k_z^4,
\label{eq:5}
\ee
where $\alpha_1 = C$, $\alpha_2 = B + C$. The frequency of columnar
fluctuations must therefore be of the form
\be
\omega_{\text{col}}({\bm k}) = \sqrt{c_z\,{\bm k}_{\perp}^2 +
c_{\perp}\,a^2\,k_z^4} .
\label{eq:1}
\ee
Here $c_z$ and $c_{\perp}$ are elastic constants, and generically $c_z \approx
c_{\perp}$ apart from a factor of $O(1)$. The application of a field $H$ that
provides a restoring force for rotations of the lattice leads to a tilt modulus
proper that is proportional to $H^2$, and hence to a term proportional to
$k_z^2$ in Eq.\ (\ref{eq:5}) whose prefactor is proportional to $H^2$.

Since the displacement vector, and hence the columnar lattice, are
two-dimensional there are two such modes; one compression mode akin to
longitudinal phonons, and one shear mode akin to transverse phonons. When the
lattice melts the transverse mode will disappear, but the longitudinal one will
survive as long as columns exist.

\begin{figure}[b]
\vskip -0mm
\includegraphics[width=8.0cm]{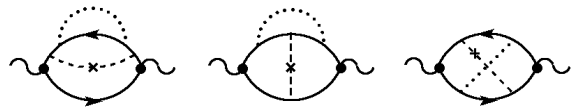}
\vskip 0mm
\caption{Leading diagrams for the conductivity in the ballistic or
 weak-disorder limit. Dotted lines represent the propagator of the columnar
 fluctuations, and crossed dashed lines represent the quenched impurities.
 The dots represent current vertices.}
\label{fig:4}
\end{figure}

Such columnar spin fluctuations will couple to the electronic charge degrees of
freedom and contribute to the electrical resistivity $\rho$, just as ordinary
magnons in antiferromagnets (which obey $\omega({\bm k}) = \sqrt{c\,{\bm
k}^2}$), contribute to $\rho$ \cite{Ueda_1977}. In a material with no
imperfections, this results in $\rho \propto T^3$, which is subleading compared
to the Fermi-liquid $T^2$ contribution. In real systems, quenched disorder is
always present. The MnSi samples where the NFL behavior is observed are clean
enough to be, for experimentally realizable temperatures, in the weak-disorder
or ballistic regime \cite{Zala_Narozhny_Aleiner_2001} where $T\tau \gg 1$, with
$\tau$ the elastic mean-free time that enters the Drude formula for the
residual ($T=0$) resistivity, $\rho_0 = \me/ne^2\tau$, with $\me$ the
electronic effective mass, $n$ the electron density, and $e$ the electron
charge. In this regime the electrical resistivity due to scattering by columnar
fluctuations can be calculated perturbatively by adapting a theory developed in
Ref.\ \onlinecite{Kirkpatrick_Belitz_Saha_2008a}. In the current case, we are
interested in scattering by excitations with a dispersion relation given by
Eq.\ (\ref{eq:1}). These are generalized phase fluctuations, so they couple to
the conduction electrons via the gradient of the fluctuation variable. The
formalism of Ref.\ \onlinecite{Kirkpatrick_Belitz_Saha_2008b} then applies, the
only difference being that the resonance frequency of the fluctuations is given
by Eq.\ (\ref{eq:1}). The leading diagrams are shown in Fig.\ \ref{fig:4}. The
conductivity tensor is anisotropic, but the temperature dependence is the same
for all components.
The result of the calculation is
\be
\delta\rho = \rho_0\,f(a\kF,\lambda/\epsilonF)\,(T/\epsilonF)^{3/2}.
\label{eq:2}
\ee
Here $\epsilonF$ is the Fermi energy, $\kF$ is the Fermi wave number, and
$\lambda$ is the energy scale associated with the magnetic order, e.g., the
Stoner splitting between electron bands that results from helical or skyrmionic
order involving one or more helices. The prefactor $f$, which depends on
dimensionless combinations of these parameters, is model dependent. It tends to
depend on $a\kF$ to a negative power. Since $a$ is given by the inverse pitch
wave number in any spin texture that involves helices, and since the latter is
small compared to $\kF$ on account of the weakness of the spin-orbit
interaction, the prefactor of the $T^{3/2}$ dependence tends to be much smaller
in model calculations than what is experimentally observed. The same is true
for calculations of the light scattering cross-section in the blue phases of
liquid crystals \cite{Lubensky_Stark_1996}. In that case, the scattering is
believed to be hugely enhanced by mode-mode coupling effects
\cite{Englert_et_al_2000, Longa_Ciesla_Trebin_2003, Ciesla_Longa_2004}, and a
similar enhancement is likely to occur in analogous magnetic systems. In a
magnetic field $H$, a term proportional to $k_z^2$ with a prefactor
proportional to $H^2$ appears under the square root in Eq.\ (\ref{eq:1}). With
parameter values appropriate for MnSi, simple considerations show that the
$T^{3/2}$ behavior will still hold for temperatures $T\agt (\mu_{\text{B}}
H)^2/k_{\text{B}} \Tc$. For fields $H \approx 0.1\,\text{T}$, see Fig.\
\ref{fig:2}, and with $T_{\text{c}} \approx 30\,\text{K}$, this yields a lower
bound for the NFL region of less than 1mK, which is consistent with the
observations.

This remarkable result suggests the following explanation of the observed NFL
behavior in MnSi. The DM interaction leads to chiral spin textures of helical
nature everywhere in the phase diagram. In the ordered phase with no magnetic
field the ground state is a single helix, which has a lower free energy than
columnar order. In the disordered phase, short-ranged columnar order (possibly
of skyrmion type, but this is not essential for our argument) is present at low
temperatures; the disordered phase is a columnar chiral fluid. It is divided
into a chiral liquid and a chiral gas; the former is the partial order phase
reported in Refs.\ \onlinecite{Pfleiderer_et_al_2004, Uemura_et_al_2007} and it
is separated from the latter by a first-order phase transition that ends in a
critical point \cite{Tewari_Belitz_Kirkpatrick_2006}. The columnar soft mode is
present in both the chiral liquid and the chiral gas as explained above. In
samples that are sufficiently clean to be in the weak-disorder or ballistic
limit, it leads to the observed NFL behavior by scattering the conduction
electrons. At temperatures higher than a few K, the short-ranged columnar order
is gradually destroyed and the NFL behavior fades. The upper temperature limit
of the ballistic regime also contributes to a crossover to a different
$T$-dependence of the resistivity. The weakness of the spin-orbit interaction
translates into a small prefactor of the $T^{3/2}$-contribution to the
resistivity in any bare theory; the quantitative observations require
anomalously large fluctuations in the disordered phase. The same is true for
the blue phases in liquid crystals, where theories yielding such an enhancement
effect have been put forward \cite{Englert_et_al_2000,
Longa_Ciesla_Trebin_2003, Ciesla_Longa_2004}. In the light of the overall
strong analogies between smectic and cholesteric liquid crystals on one hand,
and helical magnets on the other, it is plausible to expect a similar
enhancement in magnetic systems.

This work was initiated at the Aspen Center for Physics and and supported by
the National Science Foundation under Grant Nos. DMR-05-29966, DMR-05-30314,
DMR-09-29966, and DMR-09-01907.

%\bibliography{NFL}

\end{document}